\documentclass[aps,pra,showpacs,twocolumn,floatfix]{revtex4}
\usepackage{graphicx} % Include figure files
\usepackage{amsmath}
\usepackage{bm}
\usepackage{dcolumn}% Align table columns on decimal point
\begin{document}

\title{ High-accuracy calculation of black-body radiation shift
in $^{133}$Cs primary frequency standard}

\author{  K. Beloy, U. I. Safronova, and A. Derevianko }
 \affiliation{Physics Department, University of
Nevada, Reno, Nevada  89557}

\date{\today}
\begin{abstract}
Black-body radiation (BBR) shift is an important systematic
correction for the atomic frequency standards realizing the SI unit
of time. Presently, there is a controversy over the value of the BBR
shift for the primary $^{133}$Cs standard. At room temperatures the
values from various groups differ at $3 \times 10^{-15}$ level,
while the modern clocks are aiming at $10^{-16}$ accuracies. We
carry out high-precision relativistic many-body calculations of the
BBR shift. For the BBR coefficient $\beta$ at $T=300K$ we obtain
$\beta=-(1.708\pm0.006) \times 10^{-14}$, implying $6 \times
10^{-17}$ fractional uncertainty. While in accord with the most
accurate measurement, our 0.35\%-accurate value is in a substantial,
10\%, disagreement with recent semi-empirical calculations. We
identify an oversight in those calculations.
\end{abstract}

\maketitle

%\email{andrei@unr.edu}

% 2006 PACS
%32.10.Dk Electric and magnetic moments, polarizability
%06.30.Ft Time and frequency
%31.25.-v Electron correlation calculations for atoms and molecules

Since 1967, the SI unit of time, the second, is defined as a duration of a certain number of periods of
radiation corresponding to the transition between
two hyperfine levels ($F=4$ and $F=3$) of the ground state of the $^{133}$Cs atom.
In 1997, this definition has been amended, to specify that the above statement
refers to the atom at a temperature of 0 K (and at rest)~\cite{NISTSI01}.
The atomic clocks are usually operated at room temperatures and
the specific reference to $T=0$ K implies that a proper
correction for the  action of the thermal
bath of photons on the atomic energy levels is explicitly introduced.
For $^{133}$Cs clocks, it is an important
systematic correction~\cite{ItaLewWin82}, as the resulting  fractional correction to atomic frequency due to
black-body radiation (BBR) at $T=300$ is
in the order of $10^{-14}$. Moreover,
presently, there is a controversy over the value
of the BBR shift for the primary $^{133}$Cs standard. At $T=300$ K  the values
from various groups~\cite{SimLauCla98,GodCalLev05,LevCalLor04,MicGodCal04,UlzHofMor06} differ
at $3 \times 10^{-15}$ level,
while modern Cs clocks aim at $10^{-16}$ accuracies~\cite{LamAhlAsh04}.

The persistent discrepancies in the BBR shift have prompted the efforts at  the US National
Institute for Standards and Technology (NIST)
on removing the sensitivity to BBR by operating the primary standard at cryogenic
temperatures~\cite{ItanoPrivate} (BBR shift scales as $T^4$). However, because of the weight limitations,
this direct approach would be hardy feasible if
next-generation  atomic clock were to be operated at the International Space
Station~\cite{LamAhlAsh04}.
This ongoing controversy and implications for atomic time-keeping
serve as motivations for our paper. Here we compute the $^{133}$Cs BBR shift using
high-accuracy relativistic many-body techniques of atomic structure.
Our evaluated error in the BBR shift implies $6 \times 10^{-17}$ fractional
uncertainty in the clock frequency  with the value of the BBR shift consistent with
the most accurate (0.2\%-accurate) measurement~\cite{SimLauCla98}. However,
our 0.35\%-accurate value is in a substantial ~10\% disagreement with recent semi-empirical
calculations~\cite{MicGodCal04,UlzHofMor06}.
We show that this discrepancy is due to contributions of the
intermediate continuum states omitted in those calculations.

First, let us review underlying theory of the Cs BBR shift.
BBR causes a weak oscillating perturbation of atomic energy levels. Conventionally, the leading
term in the BBR contribution  is parameterized
as the fractional correction to the unperturbed clock frequency,
$\nu_0=9\, 192 \, 631 \, 770 \, \mathrm{Hz}$,
\[
  \delta \nu^\mathrm{BBR}/\nu_0 = \beta \times \left( T/T_0 \right)^4 \, ,
\]
where $T_0 = 300 \, \mathrm {K}$.
Evaluation of the coefficient $\beta$ is the goal of this work. This coefficient can be related
to the scalar differential polarizability for the hyperfine manifold of the $6s_{1/2}$ Cs ground
state.
Indeed, the characteristic
thermal photon energy at room temperatures is much smaller then the atomic energies,
so that the  perturbation
can be well described in the static limit. Moreover, contributions of electro-magnetic
BBR multipoles beyond electric dipoles, as well as retardation corrections,
are highly suppressed~\cite{PorDer06BBR}. Then the BBR shift of the energy level
is given by (atomic units are used throughout, $\alpha$ is the fine-structure
constant)
\begin{equation}
\delta E_{F}^{\mathrm{BBR}}\approx-\frac{2}{15}(\alpha\pi)^{3}\, T^{4}\alpha
_{F}(0),
\end{equation}
where $\alpha_{F}(0)$ is the static scalar electric-dipole polarizability of the hyperfine
level $F$.
The vector and tensor parts of the polarizability average out due to the
isotropic nature of the BBR.

The relation of the BBR shift to polarizability
has been exploited in the most accurate measurement of the differential Stark
shift~\cite{SimLauCla98}. However, recent direct temperature-dependent measurement~\cite{LevCalLor04}
of the BBR shift turned out to be differing by about two standard deviations from the indirect
measurement. Namely this difference has stimulated the recent interest in the Cs BBR shift.
%It seems that the only discrepancy (beyond discussed retardation and multipolar
%contributions~\cite{PorDer06BBR} and hyper-polarizability~\cite{PalDomNov03} )
% between the two types of measurements may arise
%due to dissimilarity of radiative (QED) corrections. Yet, their anticipated
%size of a few 0.1\% can
%hardly explain the discrepancy between the direct and indirect measurements.

The overall BBR shift of the clock frequency is the difference of the
individual shifts for the two hyperfine states ($F=4$ and $F=3$) involved in the transition.
While taking the difference, the traditional lowest-order polarizability of the $6s_{1/2}$ level
cancels out and one needs to evaluate the third-order $F$-dependent polarizability
$\alpha_{F}^{(3)}(0)$. This contribution involves two E1 interactions $V=- \bf{D}\cdot \bf{E}$
with the external electric field and one hyperfine $H_\mathrm{hfs}$
coupling~\cite{AngSan68}. We parameterize the general correction to the
energy  as a sum of four diagrams,
\begin{eqnarray*}
\delta E_F^{(3)} &=&
\langle F| H_\mathrm{hfs} R V R V| F \rangle+
\langle F| V R H_\mathrm{hfs} R V| F \rangle + \\
&&\langle F| V R V R H_\mathrm{hfs}| F \rangle -
\langle F| H_\mathrm{hfs}| F \rangle
\langle F| V R \, R V| F \rangle \,  ,
\end{eqnarray*}
where $R= (H-E_v)^{-1}$ is the resolvent operator, with $H$ and $E_v$
being the atomic Hamiltonian and the ground state energy, respectively.
The four contributions
will give rise to top, center, bottom, and residual
contributions discussed below. (Naming convention reflects relative
position of the hyperfine operator in the string of the three interactions). The top and bottom diagrams are equal
due to hermicity considerations. Due to angular selection rules only
magnetic dipolar interaction remains; we write
$H_\mathrm{hfs}= \bm{\mu} \cdot \mathcal{T}^{(1)}$, where $\mu$ is the nuclear magnetic moment
and $\mathcal{T}^{(1)}=-i\sqrt{2}\left( {{\bf \alpha }}%
\cdot C_{1\lambda }^{\left( 0\right) }( \hat{ r})\right) /(cr^{2})$
is the relevant relativistic coupling tensor~\cite{HFSnote}.

After angular and many-body
reduction, the  scalar
polarizability may be expressed as
\begin{eqnarray*}
\label{plrz}
\alpha^{(3)}_F(0) & = &
\frac{1}{3}\sqrt{(2I)(2I+1)(2I+2)}\left\{
\begin{array}{lll}
j_{v} &   I   & F \\
  I   & j_{v} & 1
\end{array}
\right\} \times \\
&& g_I \mu_n \left(-1\right)^{F+I+j_v}\left(2T+C+R\right)
\end{eqnarray*}
where $g_I$ is the nuclear gyromagnetic ratio, $\mu_n$ is the nuclear magneton, $I=7/2$ is
the nuclear spin, and $j_v=1/2$ is the total angular
momentum of the ground state.  The $F$-independent sums are ($|v\rangle \equiv |6s_{1/2}\rangle$)
\begin{eqnarray*}
T & = & \sum_{m,n\ne v} \frac{\left(-1\right)^{j_m+j_v}}{ 2j_v+1}
\frac{\left\langle v\left\|D\right\|m\right\rangle\left\langle
m\left\|D\right\|n\right\rangle\left\langle
n\left\| \mathcal{T}^{(1)} \right\|v\right\rangle}{(E_m-E_v)(E_n-E_v)}\delta_{j_n,j_v}\\
C & = & \sum_{m,n\ne v}\left(-1\right)^{j_m-j_n}\left\{
\begin{array}{lll}
1 & j_{v} & j_{v} \\
1 & j_{m} & j_{n}
\end{array}\right\} \times \\
& &
\frac{\left\langle v\left\|D\right\|m\right\rangle\left
\langle m\left\|\mathcal{T}^{(1)}\right\|n\right\rangle
\left\langle n\left\|D\right\|v\right\rangle}{(E_m-E_v)(E_n-E_v)} \, ,\\
R & = & \frac{\left\langle
v\left\|\mathcal{T}^{(1)} \right\|v\right\rangle}{2j_v+1}
\left(\sum_{m \in \mathrm{val} } - \sum_{m \in \mathrm{core} } \right)\frac{\left|\left\langle
v\left\|D\right\|m\right\rangle\right|^{2}}{\left(E_m-E_v\right)^{2}}.
\end{eqnarray*}
The summation indexes $m$ and $n$ range over  valence bound and continuum many-body
states and also over single-particle
core orbitals. With this convention, the above expressions subsume contributions from intermediate valence
and core-excited states and they also take into account so-called core-valence
counter-terms~\cite{DerJohSaf99}. Selection rules impose the following angular symmetries on the
intermediate states: $s_{1/2}$ for $|n\rangle$, $p_{1/2,3/2}$ for $|m\rangle$ in the top
diagram, $p_{1/2,3/2}$ for both  $|m\rangle$ and  $|n\rangle$ in the center diagram,
and, finally, $p_{1/2,3/2}$ for the $|m\rangle$ in the residual term.

We will tabulate our results in terms of the conventional scalar Stark shift coefficient
$k_s=-1/2\left( \alpha _{F=4}^{\left( 3\right) }(0)-\alpha _{F=3}^{\left( 3\right) }(0)\right) $.
For $^{133}$Cs this coefficient
can be written more explicitly in terms of the {\it F}-independent
diagrams as
\begin{equation}\label{qsc2}
k_s=-3\left(\frac{2}{3}\right)^{5/2}g_I \mu_n \left(2 T+C+R\right) \, ,
\end{equation}
with the BBR coefficient $
\beta=-4/15 \left(\alpha \pi\right)^3
T_0^4/\nu_0 \times k_s$.

Numerical evaluation of the diagrams $T$,$C$, and $R$ can be carried out either using
the Dalgarno-Lewis method or by directly summing over individual intermediate
states. Here we use the direct summation approach. This treatment is similar
to the one used in high-accuracy calculations of atomic parity violation in
$^{133}$Cs~\cite{BluJohSap92}.
 The main advantage
of this method is that one could explicitly exploit high-accuracy
experimental data for energies, dipole-matrix elements, and
hyperfine constants. When the accurate values are unknown,
we use  {\em ab initio} data of proper accuracy.
In addition, this approach  facilitates comparison with
recent calculations~\cite{MicGodCal04,UlzHofMor06}, which
also use the direct summation approach.

The central technical issue arising in direct summation over
a complete set of states is representation of the innumerable
spectrum of atomic states.
For example, even without the continuum, the bound spectrum contains
an infinite number of states. A powerful numerical method for reducing
the infinite summations/integrations to a finite number of contributions
is the basis set technique. In particular, we employ the B-spline technique~\cite{JohBluSap88}.
In this approach an atom is placed in a large spherical cavity and the single-particle
Dirac-Hartree-Fock (DHF)
orbitals are expanded in terms of a finite set of B-splines. The expansion coefficients are obtained by
invoking variational Galerkin principle. The resulting set of the single-particle orbitals is
numerically complete and finite. The technique  has a high
numerical accuracy and we refer
the reader to a review~\cite{BacCorDec01} on numerous applications of B-splines for details.

We use B-spline set with 70 splines of order 7 for each
partial wave and constrain the orbitals to a cavity of radius $R_\mathrm{cav}=
220$ a.u. This particular choice of $R_\mathrm{cav}$
ensures that the lowest-energy atomic orbitals are not perturbed
by the cavity. In particular, all core and valence DHF orbitals
with radial quantum numbers $1-12$ from the basis set produce energies and matrix elements
in a close numerical agreement with the data from traditional finite-difference
DHF code.  These low-energy orbitals will produce true many-body states for a cavity-unconstrained
atom.

To understand the relative role of various contributions, we start
by computing the Stark shift at the DHF level. We obtain:
$\frac{2T^{DHF}}{k_{S}^{DHF}}=0.418$,
$\frac{C^{DHF}}{k_{S}^{DHF}}=0.003$, and
$\frac{R^{DHF}}{k_{S}^{DHF}}=0.518$, resulting in the Stark
coefficient of $k_s^\mathrm{DHF}=-2.799\times10^{-10}$
Hz/(V/m)$^2$~\cite{numNote}. It is clear that the top and residual
terms dominate over the center diagram. The bulk (99.8\%) of the
value of the residual term is accumulated due to the principal
$6s-6p_{1/2,3/2}$ transitions.  For the top term the saturation of
the sum is not as rapid, but still the dominant contributions come
from the lowest-energy excitations: limiting the summations to the
first four excited states recovers only 68\% of the total value
for the top diagram. In addition, we find that core-excited states
contribute only 0.04\% to the final value.

The above observations
determine our strategy for more accurate calculations.
We group the entire set of atomic states into
the ``main'' low-lying-energy states (principal quantum numbers $n <=12$)
and remaining ``tail'' states. We will account for the contribution
from the ``main'' states using high-accuracy experimental
and {\em ab initio}  values. The contribution
from the ``tail'' will be obtained using either DHF or mixed approach.

First, we describe the high-accuracy data used in our calculations.
We need dipole and hyperfine matrix elements and energies.
Experimental values for the dipole matrix elements for the following
six transitions were taken from the literature (see compilations in Refs.~\cite{SafJohDer99,MicGodCal04})
$6s_{1/2}-6p_{1/2,3/2}, 7s_{1/2}-6p_{1/2,3/2},7s_{1/2}-7p_{1/2,3/2}$.
Crucial to the accuracy of the present analysis were the
matrix elements for the principal $6s_{1/2}-6p_{1/2,3/2}$
transitions. We have used 0.005\%-accurate value for $\langle 6s_{1/2} ||D|| 6p_{3/2}\rangle$
from Ref.~\cite{AmiDulGut02}. The value for $\langle 6s_{1/2} ||D|| 6p_{1/2}\rangle$ was obtained
by using the above  $6s_{1/2} - 6p_{3/2}$ matrix element and 0.03\%-accurate measured ratio~\cite{RafTan98} of
these matrix elements. These six experimental matrix elements were
supplemented by 92 values ($ns_{1/2}-n^{\prime}p_{1/2,3/2}$ values for $n,n'=6-12$ ) from high-accuracy {\em ab initio} calculations.
We employ the relativistic linearized coupled-cluster singles-doubles (LCCSD) method. The underlying
formalism, implementation, and results for alkali atoms are described in Ref.~\cite{SafJohDer99}. For dipole
matrix elements the  accuracy of the {\em ab initio} LCCSD method is a few 0.1\%.

As to the high-accuracy values of the matrix elements of the hyperfine coupling, the diagonal
matrix elements of the $\mathcal{T}^{(1)}$ tensor are directly related to the  conventional
hyperfine constants:
$A= g_I \mu_n /j_v \left[ (2j_v)/(2j_v+1)/(2j_v+2) \right]^{1/2}
\langle v || \mathcal{T}^{(1)} ||v \rangle $. We have used the
compilation of hyperfine constants from Ref.~\cite{AriIngVio77} for the ``main'' $n=6-12$ states.
Off-diagonal matrix elements between the $s$-states were evaluated
using the geometric-mean formula
\begin{eqnarray*}\label{gma}
\lefteqn{ \left|\left\langle
ns_{1/2}\left\|\mathcal{T}^{(1)}\right\|n^{\prime}s_{1/2}\right\rangle\right|
= }\\
&& \left\{ \left|
\langle ns_{1/2}\left\|\mathcal{T}^{(1)}\right\|ns_{1/2}\rangle
\langle
n^{\prime}s_{1/2}\left\|\mathcal{T}^{(1)}\right\|n^{\prime}s_{1/2}\rangle\right|
\right\}^{1/2}.
\end{eqnarray*}
This expression  has been shown to hold to about $10^{-3}$ in Ref.~\cite{DzuFla00} (notice that the radiative
corrections would start playing a role at a few 0.1\% as well).
If we had $6\le n \le 12$ in the above expression, then the experimental value
is used for its corresponding diagonal element on the right. If we
also had $6 \le n^{\prime} \le 12$, then that experimental value is also used
for the corresponding diagonal element on the right, otherwise the
DHF value is taken. ($n$ and $n'$ can be obviously interchanged in this prescription.) This mixed approach has allowed us to uniformly improve the accuracy
of the calculations. Indeed, in the numerically
important top term, the hyperfine matrix elements
come in the combination
$\langle ns_{1/2}\left\|\mathcal{T}^{(1)}\right\|6s_{1/2}\rangle$.
As $n$ grows, the correlations become less important, so the dominant
correlation correction comes from the $6s_{1/2}$ state. Using the described
mixed approach allows us to account for this dominant correlation.
The geometric-mean formula holds only for the $s$ states.
For the off-diagonal matrix elements  between various combinations of
$6p_{1/2,3/2}$ and $np_{1/2,3/2}$ ($n=6-9$) states we employed
a modification of the LCCSD method augmented by perturbative
treatment of the valence triple excitations (LCCSDpvT method)~\cite{SafJohDer99}.
The accuracy of these matrix elements is a few \%. As these matrix
elements enter the relatively small center term, the effect on the
overall theoretical error is negligible.

Finally,  we used
experimental energy values from the NIST tabulation~\cite{Moo58},
for states with principle quantum number $n=6-12$, and the DHF
values otherwise.

%Replacing these Dirac-Fock matrix elements and energy values results
%in a scalar Stark coefficient of
%$k=-2.268\times10^{-10}$ Hz/(V/m)$^2$.  This corresponds to a
%relative blackbody radiation shift of $\beta=-1.708\times10^{-14}$
%at 300K.

With the described set, we report
our final result for the scalar Stark coefficient and relative
blackbody radiation shift at 300K to be
\begin{eqnarray}
k_s  &=& -(2.268\pm0.008)   \times 10^{-10} \, \, \mathrm{Hz/(V/m)^2} \, , \nonumber \\
\beta &=& -(1.708 \pm 0.006) \times 10^{-14} \,. \label{Eq:final}
\end{eqnarray}
The values of the individual diagrams are
$\frac{2T}{k_{S}}=0.449$, $\frac{C}{k_{S}}=-0.002$, and
$\frac{R}{k_{S}}=0.553$. When comparing with the DHF values, the
most substantial modification due to correlations is in the center
term, which changes the sign. Fortunately, this term is relatively
small, and this extreme change does not substantially  affect the
final result.

The overall uncertainty of these results was determined from the
uncertainties of the individual matrix elements and energy values
used in their computation.  Standard uncertainty analysis was done
throughout all mathematical operations.  For energy values taken
from NIST, the uncertainty is assumed negligible.  For all other
experimental values, the reported uncertainty is used.  For {\em ab
initio} matrix elements (DHF, LCCSD, or LCCSDpvT) we assigned an
assumed uncertainty.  These assumed uncertainties were based on
comparison between calculated and high-accuracy experimental values.
This resulted in a relative uncertainty for both the scalar Stark
coefficient and the BBR shift of 0.35\%. We have performed several
consistency checks, e.g., replacing experimental matrix elements and
energies by {\em ab initio} LCCSD values or by replacing the DHF
values for states with $n=13-27$ with the LCCSD values. The final
result was stable to such modifications within the stated
uncertainty in Eq.(\ref{Eq:final}). These tests provide us with
additional confidence with respect to our standard error analysis
based on errors of used experimental values. It is also worth noting
that the present calculation does not include radiative corrections
which may contribute at the level of a few 0.1\% (some radiative
corrections, e.g., vacuum polarization, are absorbed in our final
value already as we use {\em experimental} hyperfine constants).

A comparison with  recent theoretical and experimental work is
presented in Table~\ref{Tab:ksComp}. While agreeing with the most
accurate measurement by \citet{SimLauCla98}, our results are in
substantial disagreement with the recent calculations
\cite{MicGodCal04,UlzHofMor06}. The principal
differences between the present work and these calculations are: (i)
more sophisticated treatment of correlations, and (ii) rigorous
summation over the {\em complete} set of intermediate states in
perturbative expressions. As discussed above, we used the numerically
complete basis-set approach which approximates  Rydberg states
and  continuum with quasi-spectrum. By contrast, in
Ref.~\cite{MicGodCal04}, the summations were truncated at $n=9$, and
in Ref.~\cite{UlzHofMor06} at $n=18$; neither  work includes
continuum. To illuminate the importance of the omitted contributions
we truncate our summations at $n=12$. The resulting value deviates
from our final $k_s$ result by $0.29 \times 10^{-10}$ Hz/(V/m)$^2$.
This large 10\% ``continuum correction'' brings the values from
Refs.~\cite{MicGodCal04,UlzHofMor06} into essential agreement with
our result. The fact that continuum needs to be included is hardly
surprising, as, for example, about 20\% of the textbook
polarizability of the ground state of the hydrogen atom comes from
the continuum states.

%%We present the comparison with the previous
%%These numbers disagree with the theoretical results
%$k=-(1.97\pm0.09)\times10^{-10}$ Hz/(V/m)$^2$ and
%$\beta=-(1.49\pm0.07)\times10^{-14}$ given by Micalizio {\it et al.}
%in Ref.~\cite{MicGodCal04}.  Our numbers also disagree with the
%experimental results $k=-(2.05\pm0.04)\times10^{-10}$ Hz/(V/m)$^2$
%given by Godone {\it et al.} in Ref.~\cite{GodCalLev05} and
%$\beta=-(1.43\pm0.09)\times10^{-14}$ given by Levi {\it et al.} in
%Ref.~\cite{LevCalLor04}.  However, our results agree to an excellent
%degree with the experimental results
%$k=-(2.271\pm0.004)\times10^{-10}$ Hz/(V/m)$^2$ given by Simon {\it
%et al.} in Ref.~\cite{SimLauCla98} and
%$\beta=-(1.66\pm0.20)\times10^{-14}$ given by Bauch and Schr\"{o}der
%in Ref.~\cite{BauSch97}.

\begin{table}
\caption{\label{Tab:ksComp} Values of $k_S$ in 10$^{-10}\mathrm{Hz/(V/m)}^{2}$.}
\begin{ruledtabular}
\begin{tabular}{lll}
\multicolumn{1}{c}{ }&
\multicolumn{1}{c}{ }&
\multicolumn{1}{c}{References}\\
\hline
theory& -1.97$\pm$ 0.09&  Ref.~\cite{MicGodCal04}\\
theory& -2.06$\pm$ 0.01&Ref.~\cite{UlzHofMor06}\\
expt.&-2.05$\pm$ 0.04 & Ref.~\cite{GodCalLev05}\\
expt.&-2.271$\pm$ 0.004&  Ref.~\cite{SimLauCla98}\\[2ex]
theory &-2.268$\pm$ 0.008&   present           \\
\end{tabular}
\end{ruledtabular}
\end{table}

To conclude, here we have reported results of relativistic many-body
calculations of the BBR shift, one of the leading systematic correction in $^{133}$Cs frequency
standard and a subject of the recent controversy.
Our 0.35\%-accurate result re-validates high-precision Stark shift measurements
\cite{SimLauCla98}. Our work also clarifies the origin of the reported
discrepancy between that measurement and recent
calculations~\cite{MicGodCal04,UlzHofMor06}.

We would like to thank H. Gould, M.-A. Bouchiat, and W. Itano for
discussions. Work of K. B and A.D was supported in part by NIST
Precision Measurement Grant, and National Science Foundation. Work
of U.I.S was supported in part by
 DOE-NNSA/NV Cooperative Agreement DE-FC52-01NV14050.

{\em Note.} While completing writing this manuscript, we have
learned of another accurate many-body calculation of the BBR shift
in $^{133}$Cs clock~\cite{AngDzuFla06}. Their result, $k_s= 2.26
\times 10^{-10} \pm 1\% \, \mathrm{Hz/(V/m)^2}$, is in agreement
with our more accurate value.

%\bibliography{all,CsBBRnotes}

\begin{thebibliography}{24}
\bibitem[{HFS()}]{HFSnote}
\bibinfo{note}{This coupling is for a point-like nucleus. In calculations we
  include the effect of finite nuclear magnetization, modeling the nucleus as a
  uniformly magnetized ball of radius 5.6748 fm.}
\bibitem[{num()}]{numNote}
\bibinfo{note}{To test the numerical convergence, we also used another basis
  set of 40 splines with $R_\mathrm{cav}= 75$ a.u. The resulting number was
  consistent with the value produced with the described larger set to one part
  in 40,000.}
\end{thebibliography}

\end{document}